%% file: ABManu_Ahmed24-final.tex
\documentclass[
 aps, pra,
 amsmath,amssymb,
 11pt,
 final,
tightenlines,
 twoside,
 twocolumn,
 nofloats,
nofootinbib,
 superscriptaddress,
showkeys,
showkeywords,
 ]
{revtex4-2}

\usepackage[T1]{fontenc}
\usepackage[utf8]{inputenc}
\usepackage[english]{babel}
\usepackage{graphicx,subcaption}% Include figure files
\usepackage{dcolumn}% Align table columns on decimal point
\usepackage{bm}% bold math

\input{maik.rty}

\setcitestyle{authoryear,round}
\input{sao_cmd_author.tex}

\begin{document}
\nocite{*}
\selectlanguage{english}


%\ydk{}
%\titlerunning{}
%\authorrunning{}
%\toctitle{}
%\tocauthor{}

\title{Deep Photometric and Astrometric Investigation of the Non-relaxed Star Cluster Stock 3 using Gaia DR3}

\author{A. ~\surname{Ahmed}}
 \email{ahamza@sci.cu.edu.eg}
 \affiliation{Astronomy, Space science and Meteorology Dept., Faculty of Science, Cairo University}

\author{Amira ~\surname{R. Youssef }}
 \affiliation{Astronomy, Space science and Meteorology Dept., Faculty of Science, Cairo University}
 
\author{M.S. ~\surname{ El-Nawawy}}
 \affiliation{Astronomy, Space science and Meteorology Dept., Faculty of Science, Cairo University}

 \author{W. H. ~\surname{ Elsanhoury}}
 \affiliation{Physics Department, College of Science, Northern Border University, Arar, Saudi Arabia}
  \affiliation{ Astronomy Department, National Research Institute of Astronomy and Geophysics (NRIAG), Helwan, Cairo, Egypt}
 
\begin{abstract}
The study presents both photometric and kinematic analyses of the non-relaxed open cluster Stock 3 with Gaia DR3 which found to be positioned at  2.945 $\pm$ 0.700 kpc and having an age of 16.00 $\pm$ 4.00 Myr. We analyse the data to infer the membership and thus determine the total mass, IMF and the dynamical and kinematical status.

\end{abstract}

\maketitle

\section{INTRODUCTION}

The deep analysis of the open star clusters significantly enhances our understanding of the stellar evolution. Additionally, they widen our knowledge about the structure and evolution of the Milky Way (MW) thin disc \citep{gilmore2012gaia}. Stock 3, which located at Cassiopeia constellation in the northern MW, is part of the 21 open clusters first observed by the German Astronomer Jürgen Stock \citep{macconnell12006homage} in the early 1950s at the Cerro Tololo observatory site in Chile \citep{tonkin2013binocular}. Several extensive catalogues have been compiled, providing derived parameters for this cluster, for instance, \cite{dias2002new,kharchenko2013global,dias2014proper,sampedro2017multimembership}. The results of many of the latest studies that investigated the cluster is mentioned in Table \ref{Stock3_results} below.

%\subsection{Stock 3}
According to Webda database, the celestial coordinates of Stock 3 is positioned at Right Ascension (RA) ($\alpha$ $= 01^{h} 12^{m} 18^{s}.00$) and Declination (Dec.) ($\delta$ $= 62^{\circ} 20' 00''.0 $), corresponding  to  the Galactic coordinates  $\ell = 125^{\circ}.351$ and $b= -0^{\circ}.550$. \cite{kharchenko2013global} presented the parameters of Stock 3 in their MWSC catalogue that is based on the 2MASS and PPMXL databases.
Using the same catalogue, \cite{joshi2016study} calculated the Galactic coordinates for Stock 3. While using the UCAC4 catalogue, \cite{sampedro2017multimembership} derived the proper motion components of Stock 3. Also, \cite{monteiro2020fundamental} analysed Stock 3 using Gaia DR2. The results of these studies are outlined in Table \ref{Stock3_results}.
%
% %%%%%%%%%%%%%%%%%
%  RESULTS OF Stock   3  PLUS COMPARISON
%\begin{landscape}
\begin{table*}
\caption{Comparison of the estimated parameters in the literature:($\alpha$, $ \delta$; deg.), radius (r; arcmin), age (Myr), distance (d; pc), colour excess $E(B - V)$; mag, parallax ($\varpi$; mas), proper motion components ($\mu_\alpha \cos \delta, \mu_\delta$; mas yr\textsuperscript{-1}), the number of member stars and the references.
\label{Stock3_results}}    
\setlength\tabcolsep{3pt}
\scalebox{0.9}{
\begin{tabular}{cccccccccccc}
\hline
$\alpha $ &  $\delta$  &  r & Age   & d &\it E(B-V) & $ \varpi$ &  $ \mu _\alpha \cos \delta$ & $ \mu _\delta$  & N & Ref.  \\
\ deg & deg & $ arcmin $ & Myr & pc & mag & mas &  mas $\rm yr^{-1}$ &mas $\rm yr^{-1}$& stars & \\
\hline
%18.040 & 62.256 &5.60 & 16.00$\pm$4.00 & 3203.1 &0.97& 0.3122$\pm$0.0351 &-2.0673$\pm$0.0173& -0.4657$\pm$0.0192  &73 & 1 \\
18.075  & 62.334  & 3.50  &--- & --&--&--& $-$1.94$\pm$3.73 &  1.54$\pm$2.49&59&1 \\ 
18.038  & 62.258  &---  & 70.79 & 1363.0&0.708&--&--& --&  --& 2 \\
18.037 &62.258  & 6.60  & 70.79& 1363.0&0.708&--& $-$0.13 & $-$0.64&90&3\\
18.060 & 62.319 & --- & 16.80$\pm$3.40  &2747.0 &--&0.265$\pm$0.132 & $-$1.895$\pm$0.326  &$-$0.357$\pm$0.296&114&4 \\
18.075 & 62.334 & 4.00  & 70.79 & 1363.0&0.71 &--& $-$1.96$\pm$0.62  & 1.68$\pm$0.5 &64&5 \\
\hline \\
\end{tabular}}
References: (1)\cite{dias2014proper}; (2) \cite{joshi2016study};  (3) \cite{kharchenko2013global}; (4)  \cite{monteiro2020fundamental}; (5) \cite{sampedro2017multimembership}.
\end{table*}
%  d($m-M$) (kpc) &  2.7103 $\pm$ 0.70 & 2.6023 $\pm$ 0.66  \\%  &&\\

As shown in the table \ref{Stock3_results}, there are huge discrepancies and lack of similarities between the estimated astrophysical parameters. This might be attributed to the usage of different datasets, observational techniques, and different statistical approaches for the selection of members.  
This agrees with the findings of \cite{netopil2015comparative} who made an extensive comparison of the results of many previous studies that provided estimates of the basic parameters for large numbers of open clusters, e.g. \cite{kharchenko2013global,Bukoetal11}. They compared the measured ages, distances and colour excesses of the mutual open star clusters and found out that there are non-uniformities in their estimations. We aim to do more refinement of the results of the considered open cluster using the most recent data set, Gaia DR3.
The paper is coordinated as follows. The data used is described in the next section. 
Measured centre and radius are presented in Section \ref{centres-radii-sect}. The selection of the probable members from Gaia DR3 database in addition to their proper motions and parallaxes are discussed in Sections \ref{gaia-membership} and \ref{Kin-prop}, respectively.  Evaluations of the age and the distance of the cluster are reported in Section \ref{CMD-Age-Dist}. The estimated luminosity and mass function with the dynamical status of the cluster are addressed in Sections \ref{LM-MF-sect} and \ref{dyan_state_sect}, respectively. Section \ref{VEP_sec} is devoted to the velocity ellipsoid parameters.At the end, we present conclusions in Section \ref{conc-sect}.

\section{ Data Used}
\label{data-sect}

In the present work, we have used Gaia DR3\footnote{\url{https:://www.cosmos.esa.int/gaia}} \citep{arenou2022gaia, bailer2022gaia, vallenari2023gaia, brown2021gaia, lindegren2021gaia,prusti2016gaia}  to get the astrometric and photometric data of Stock 3.

A new chapter in Astronomy has begun since the launch of the European Space Agency mission Gaia as it contains the five-parameter astrometry for approximately 1.80 billion sources along with their positions on the sky ($\alpha$,$\delta$), parallaxes ($\varpi$) and the RA and Dec. components of the proper motion ($\mu_{\alpha}\cos\delta$, $\mu_{\delta}$) with a limiting magnitude of $G$= 21 mag. The uncertainties in the respective proper motion components are up to 0.02-0.03 mas $\rm yr^{-1}$ (at $G<$ 15 mag), 0.07 mas $\rm yr^{-1}$ (at $G\sim$ 17 mag), 0.50 mas $\rm yr^{-1}$ (at $G\sim$ 20 mag) and 1.40 mas $\rm yr^{-1}$ (at $G$ = 21). The uncertainties in the parallax values are$\sim$ 0.02–0.03 mas for sources with $G<$15 mag,$\sim$ 0.07 mas for sources with $G$=17 mag,$\sim$ 0.50 mas at $G$= 20 mag and$\sim$ 1.30 mas at $G$= 21. 
Running analyses and visualizations with the computer program TOPCAT\footnote{\url{http://www.starlink.ac.uk/topcat/}}, we selected the raw data from Gaia DR3 by specifying a $\rm 10'$ radius area around the central coordinates of the cluster, in such a way that to exceed the reported radius of it. Gaia DR3 is exhibiting much enhancement over Gaia DR2; precision of parallax measurements has improved by 30 \%, and the accuracy of the proper motion measurements has improved by a factor of 2.00. 

\section{Cluster's Center and Radius}\label{centres-radii-sect}

Cluster's centre can be determined by locating the area with the highest star density. To achieve this, we constructed histograms of $\alpha$ and $\delta$, dividing the extracted region into bins of identical size and applying Gaussian fitting, the cluster’s centre exits at $\alpha =01^{h} 12^{m} 9^{s}.40$ and $\delta= +62^{\circ} 15 '23.08 ''$; which concurs with the literature. 
\begin{figure}%[htb!]
\captionsetup[subfigure]{labelformat=empty} % to remove subfigure labels
%\centering
%\hspace{-3.0cm}
\subfloat[]{\scalebox{.52}{\includegraphics{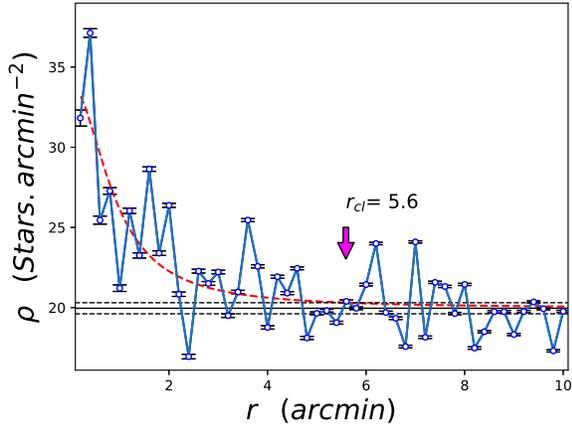}}}
\caption{The RDP represented by the blue solid line. The dashed red line represents the \cite{king1962structure} fitting while the black solid line and  dashed line mark the background density $f_{bg}$ and the uncertainties of the the background density, respectively. 
%The length of the error-bars denote errors resulting from sampling statistics, in accordance with Poisson distribution ($= 1/\sqrt{N}$, where N is the number of stars used in the density estimation at that point).
\label{rdps}}
\end{figure}

To study the structure of the cluster, we drew the radial density profile (RDP), as shown in Figure \ref{rdps},  by dividing the observed area into concentric shells from the cluster centre with equal steps of 0.20 arcmin. Each zone's density ($\rho_i$; stars arcmin\textsuperscript{-1})  was computed by dividing the number of stars within it by its area, i.e. $\rho_i=\frac{N_i}{A_i}$.
The uncertainties in the density calculatiions were determined from sampling statistics, in accordance with Poisson distribution ($= 1/\sqrt{N_i}$).
 The radius of the cluster is defined as the distance at which the star density matches that of the field star density \cite{tadross2005analytical}. Applying the empirical King’s Model \citep{king1962structure}, the density function $\rho_r$ can be represented as:

\begin {equation}
\rho_r= f_{bg} + \frac{f_0}{1+ (r/r_c)^2}\,, 
\end {equation}
where $f_{bg}$, $f_0$ and $r_c$ are background density, central star density, and the core radius (which is the distance where the stellar density drops to half of the central density) of the cluster, respectively. The best fit model is the model with the highest coefficient of determination 
\begin{equation}
R^2=1-\frac{SS_{res}}{SS_{tot}}\,,
\end{equation}
 $SS_{res}$ is the residual sum of squares computed by $SS_{res}=\sum_i(y_i-f_i)^2$ where $y_i$ represents the observed RDP, $f_i$ represents the King's profile, and $SS_{tot}$ is the total sum of squares computed by $SS_{tot}=\sum_i(y_i-\overline{y})^2$, where $\overline{y}$ is the mean of the observed data, $\overline{y}=\frac{1}{n}\times\sum_{i=1}^n y_i$.

The limiting radius of Stock 3 is found to be 5.60 $\pm$ 0.43 arcmin, which is not too far from \cite{kharchenko2013global}, $f_{bg}$= 19.93 $\pm$ 0.35 stars $\rm arcmin^{-2}$, and $ r_c$=0.79 $\pm$ 0.13 arcmin.
As shown in the Table \ref{Stock3_results}, the radii estimates obtained using IR data are distinguishably larger than the calculations using the optical data \citep{joshi2015basic}.
%------------------------------------------------------------------------------------------------------------------------------------------------
%------------------------------------------------------------------------------------------------------------------------------------------------
%
%\begin{figure*}[htp]
\begin{figure}[htp]
\captionsetup[subfigure]{labelformat=empty} % to remove subfigure labels
%\centering
\subfloat[]{\scalebox{.42}{\includegraphics{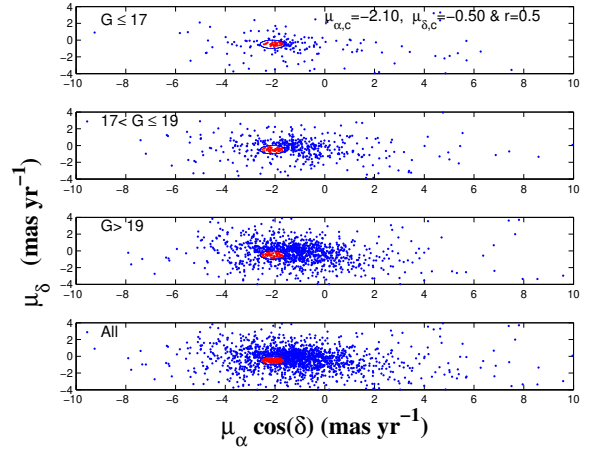}}}\hfil

\caption{VPD of the stars in the specified G magnitudes that lay in the estimated radius of the cluster. Red points represent stars lay within the selected cluster zone.
\label{VPDs_GMag}}
\end{figure}

\section{Membership determination using Gaia DR3}\label{gaia-membership}

By making use of the kinematic data, The next crucial step in determining all the astrophysical parameters is to deduce the cluster members with the highest probabilities. The proper motion approach is favoured over radial velocities as the former draws movements in two dimensions instead of only one as in the latter \citep{tian1998determination}. %
Membership was determined following the statistical approach of \cite{yadav2013proper} which is based on the procedure of \cite{balaguer1998determination}. 
\cite{yadav2013proper} procedure starts by identification of the cluster zone on the vector point diagram (VPD), which is achieved by plotting proper motion components ($\mu_\alpha \cos\delta$, $\mu_\delta$)  at different intervals of the {\it G} magnitudes, as shown in Figure \ref{VPDs_GMag}. The region of the highest stellar population compared to the rest of the field is considered as the cluster domain which is clearly seen in the {\it $G\leq$} 17 and 17 {\it $<G\leq$} 19 intervals. This region is 0.50 mas $\rm yr^{-1}$ radius centred at $\mu_\alpha \cos \delta$= $-$2.10 and $\mu_\delta$= $-$0.50  mas $\rm yr^{-1}$. The stars that exist outside this circular region were treated as field stars.
The next step is to compute the membership probabilities of the stars that lie within the estimated radius of the cluster. We began by calculating the frequency distribution for the cluster’s stars ($\phi_c^\nu$ ) and for the field stars ($\phi_f^\nu$) respectively, using equations 3 \& 4 of \cite{balaguer1998determination}.
In these equations, the intrinsic proper motion dispersion of cluster member stars was given a value of 0.075 mas $\rm yr^{-1}$ which relies on the assumption of \cite{girard1989relative} that the radial velocity dispersions within open clusters are estimated to be 1 km $\rm s^{-1}$ \citep{sinha2020variable} . 
We found that the average of the proper motions components of all field stars, and their corresponding dispersions are $\mu_{xf}$ = $-$0.66 mas $yr^{-1}$, $\mu_{yf}$ = $-$0.42 mas $yr^{-1}$, $\sigma_{xf}$ = 4.25 mas $yr^{-1}$ and $\sigma_{yf}$= 2.05 mas $yr^{-1}$.

The distribution of all stars can be derived from

\begin {equation}
\Phi = \left(n_{c}\cdot\phi_{c}^{\nu}\right) +  \left(n_{f}\cdot\phi_{f}^{\nu}\right)\,,  
\end {equation}
where $n_c$ and $n_f$ denote the normalized number of cluster and field stars, respectively, which are equal to 0.08 and 0.92, keeping in mind that  $n_c + n_f= 1$.
Accordingly, the probability of $\rm i^{th}$ star’s membership is given by:

\begin {equation}
P_{\mu}(i) = \frac{\phi_{c}(i)}{\phi(i)}\,.  \end {equation}

\begin{figure}[htp]
\centering
\subfloat[]{\scalebox{0.4}{\includegraphics{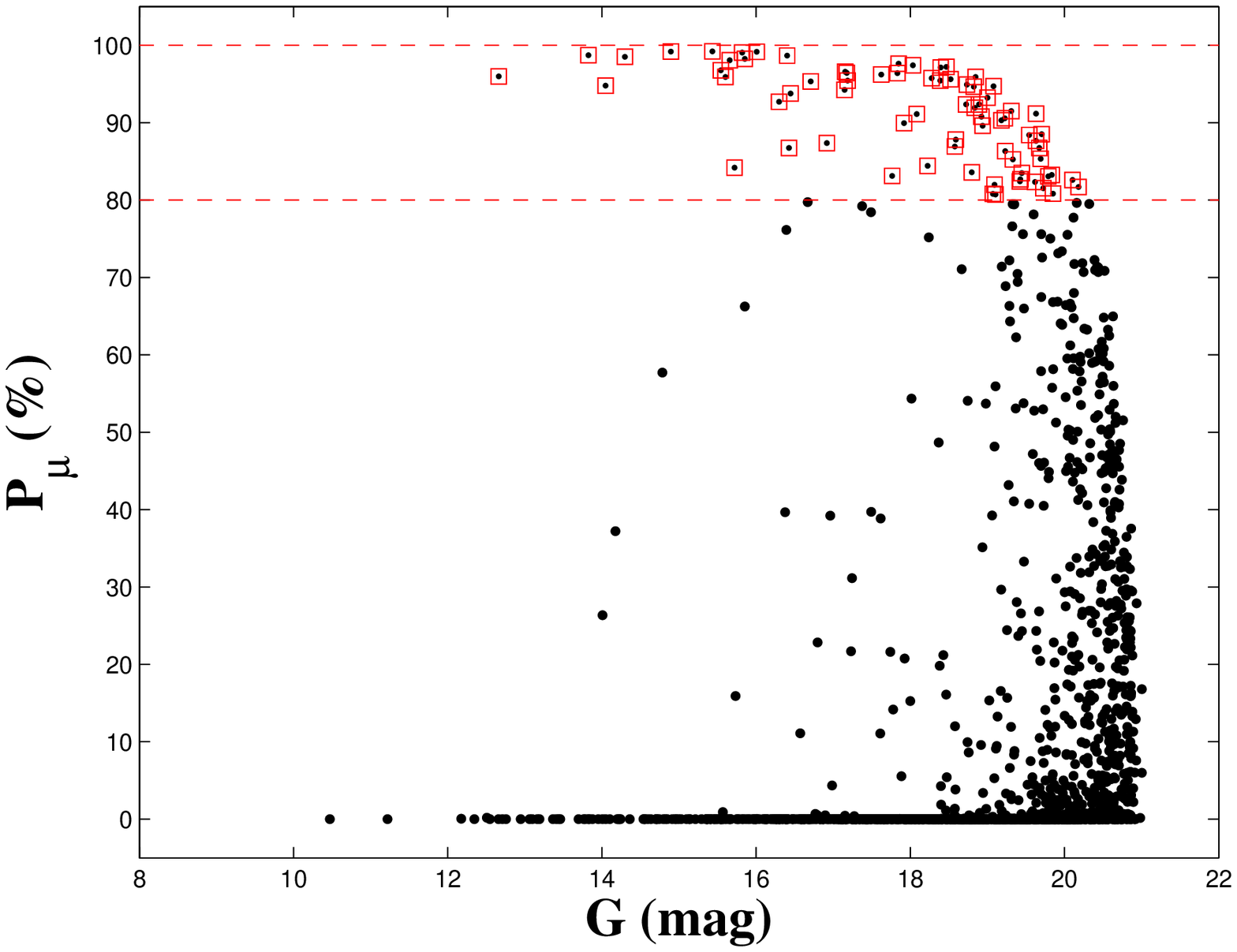}}\label{3A}}

\subfloat[]{\scalebox{.4}{\includegraphics{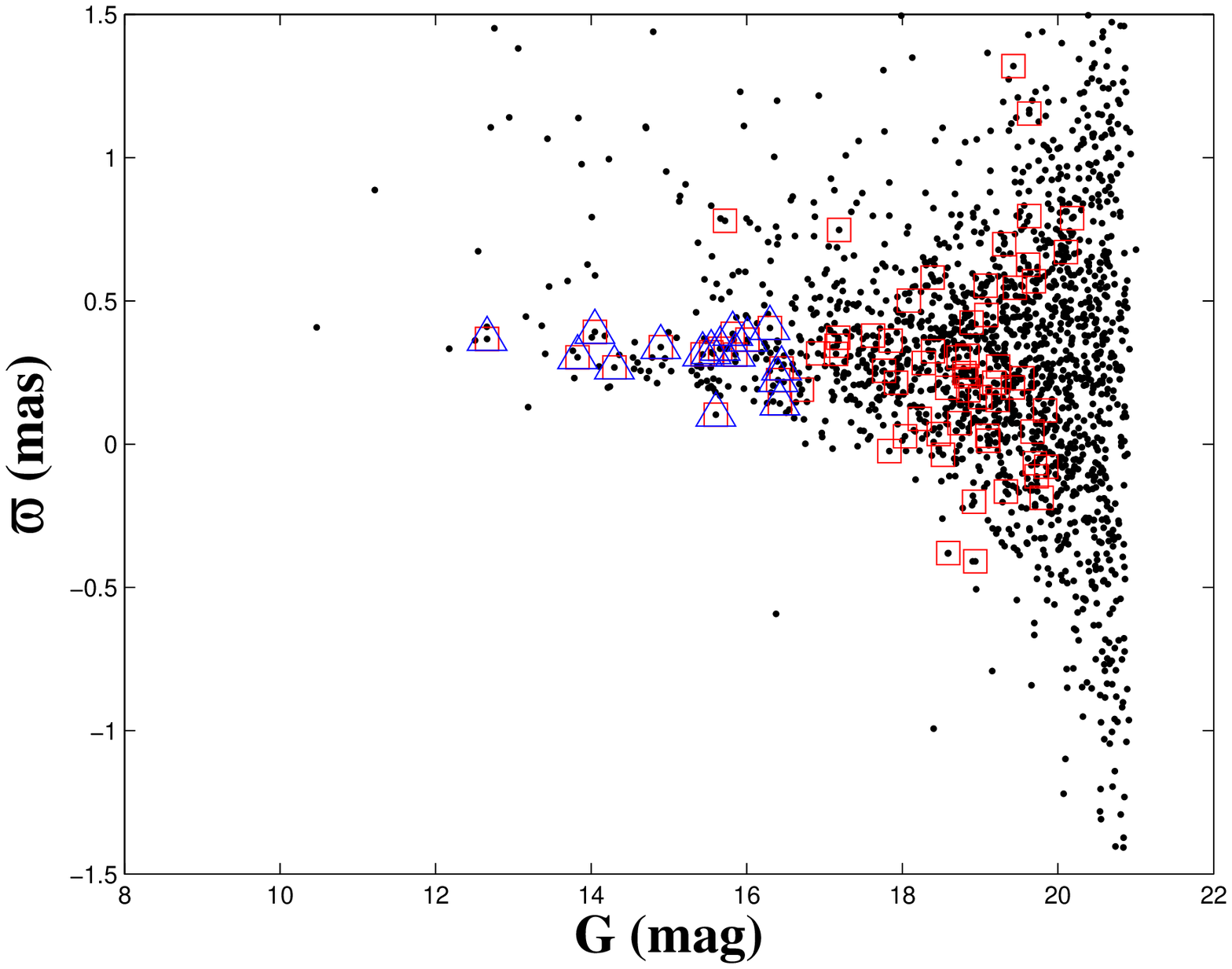}} \label{3B}}

\caption{Plots of membership probability ($P_\mu$) (a) and parallax ($\varpi$) (b) as a function of $G$ mag, respectively. The black points represent all stars while the red squares in both panels mark the stars with membership probability $\geq$ 80\% ( which is at or above the horizontal line) . Blue triangles in the parallax figure label the members of parallax errors $\leq$ 0.05 mas.
\label{Prob_Plx}}
\end{figure}

\begin{figure}[htp]
\captionsetup[subfigure]{labelformat=empty} % to remove subfigure labels
\centering
\subfloat[]{\scalebox{.5}{\includegraphics{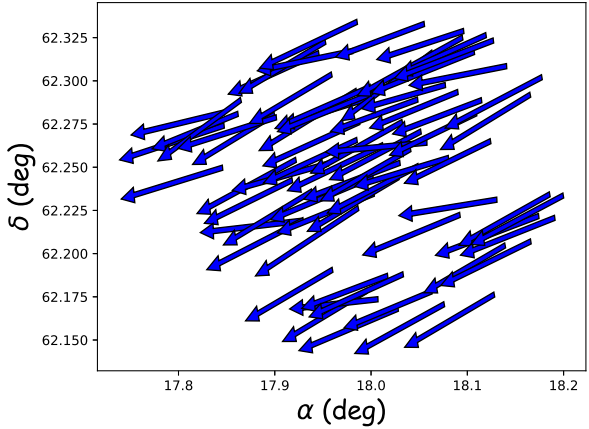}}}
\caption{Co-moving stars, where blue arrows represent stars with probability $\geq$ 80\%. %while stars with probability $\geq$ 95\% are represented by red arrows.
\label{comoving_vector}}
\end{figure}

The membership probability and parallax are plotted as a function of $G$ (mag), as presented in Figure \ref{Prob_Plx}. In the figure, stars with membership probabilities equal to or higher than 80\% are marked with red squares, while the stars with blue triangles in  panel (\subref{3B}) represent stars with highly precise measurements of parallax, $\varpi_{err}\leq0.05$ mas. 
In present study, we have found 73 members with probabilities $\geq$ 80\% down to a {\it G} magnitude of 20.2 mag as shown in Figure \ref{Prob_Plx}. 
In panel (\subref{3B}) of Figure \ref{Prob_Plx}, the parallax estimates of the probable members marked by red squares have a smaller dispersion compared to those of the field stars. Also,  it shows that members with highly accurate parallax measurements marked by blue triangles have even a much smaller dispersion. 
By using the measured components of the proper motions of the probable members,  we drew the co-moving stars,  as shown in Figure \ref{comoving_vector}; it demonstrates uniformity in direction and equality in speed.

To determine the effectiveness of the membership determination, we used the following formula of \cite{shao1996effectivity}
\begin {equation}
E=1-\frac{N \sum_{i=1}^{N} [P(i)(1-P(i))]}{\sum_{i=1}^{N} P(i) \sum_{i=1}^{N} (1-P(i))}
 \, , \end {equation}
where higher values of $E$ reflect high effectiveness of the procedure. The study of \cite{shao1996effectivity} shows that $E$ varies between 0.2 to 0.9 and it has an optimum value at 0.55. The effectiveness of membership determination is found to be 0.54.  This demonstrates that the effectiveness of members selection is significantly high.
\begin{figure}[htp]
%\captionsetup[subfigure]{labelformat=empty} % to remove subfigure labels
\centering
\subfloat[]{\scalebox{0.4}{\includegraphics{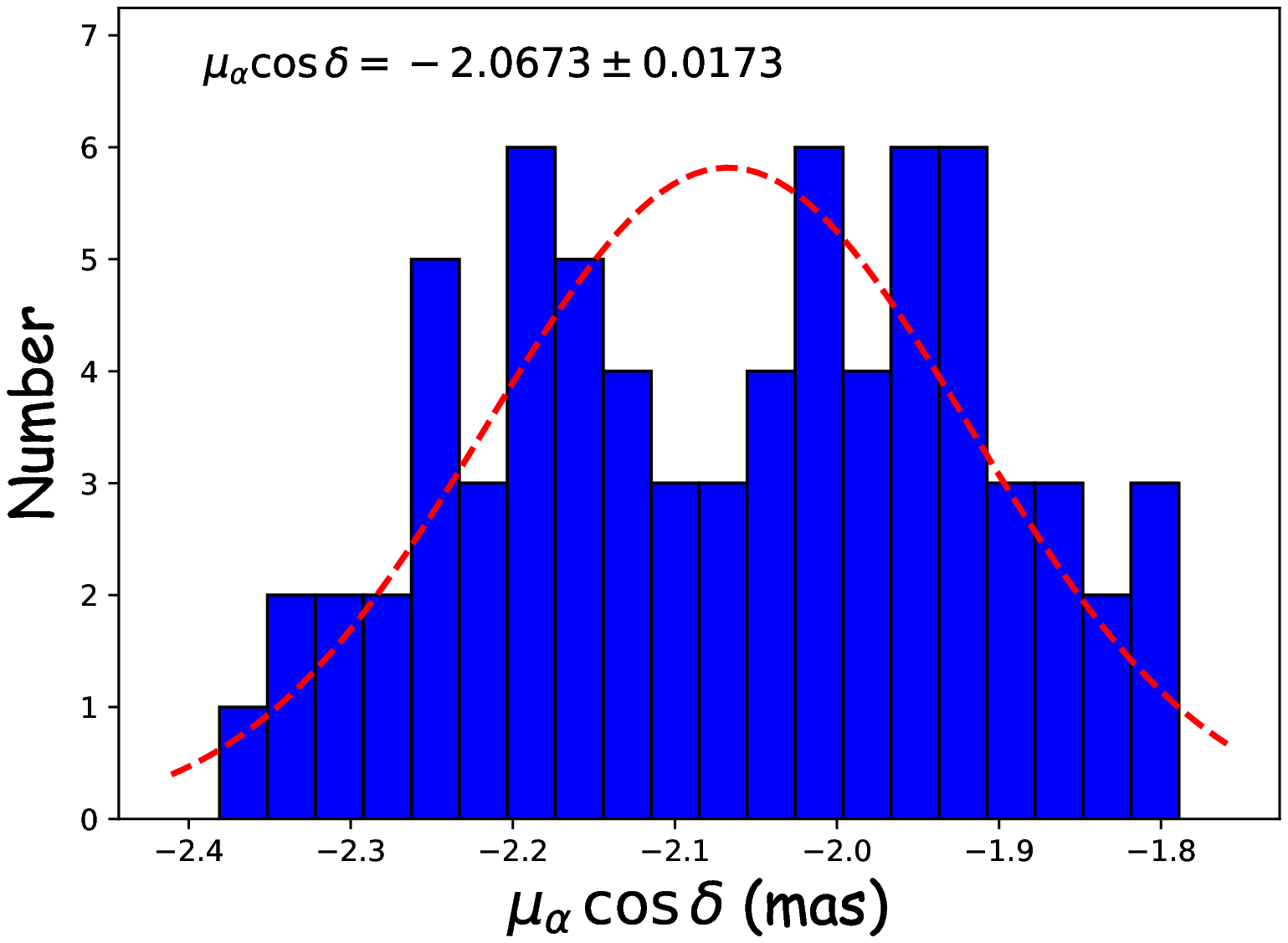}}}\\
\subfloat[]{\scalebox{0.4}{\includegraphics{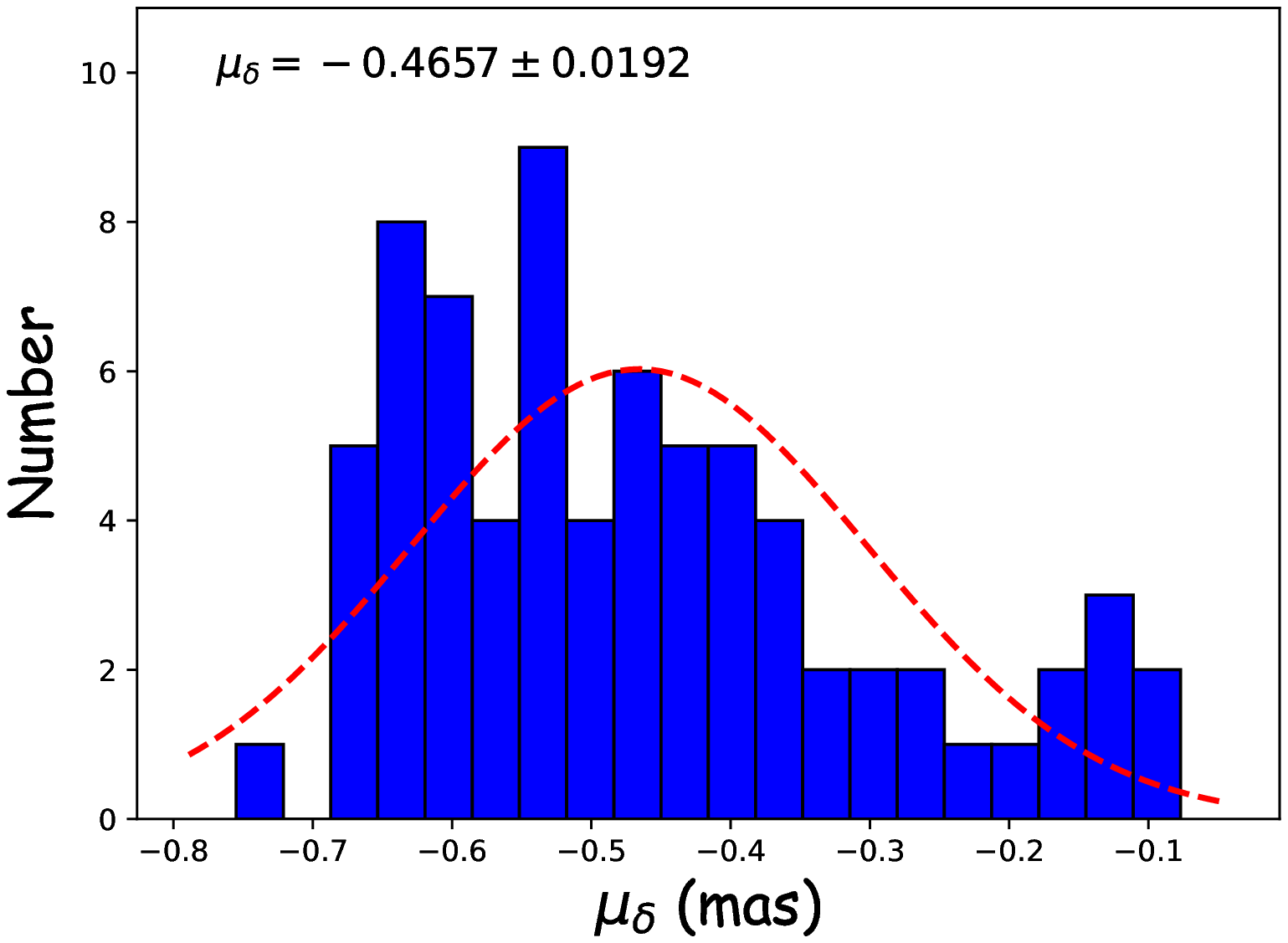}}}

\caption{Proper motion ranges in RA and Dec. The Gaussian fitting of the bins represented by red lines in the panels. 
\label{PM_Hists}}
\end{figure}
\section{Kinematic Properties of the Cluster}\label{Kin-prop}

The proper motion and parallax of the cluster were computed using the data of the selected members. 
We constructed histograms of the astrometric data of the 80\% probable members and by applying Gaussian fitting to these histograms; the mean proper motion components of Stock 3 are $-$2.0673 $\pm$ 0.0173 mas $\rm yr^{-1}$ and $-$0.4657 $\pm$ 0.0192 mas $\rm yr^{-1}$ respectively, in right ascension and declination as illustrated in Figure \ref{PM_Hists}. 
Our estimates of the proper motion components matches well with the calculations of \cite{monteiro2020fundamental}.
Also, ours results agree with those of \cite{dias2014proper} and \cite{sampedro2017multimembership} who used different datasets, but they have higher uncertainties. On the contrary, our values differ significantly from those of \cite{kharchenko2013global}, because they used the PPMXL database which have much less precise proper motion measurements compared to either Gaia DR3 or Gaia DR2.
\begin{figure}%[htp]
\captionsetup[subfigure]{labelformat=empty} % to remove subfigure labels
\centering
\subfloat[]{\scalebox{0.4}{\includegraphics{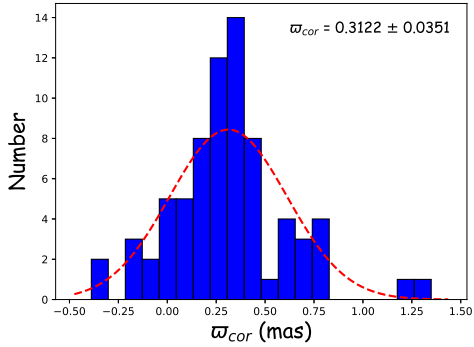}}} 

\caption{The parallax histogram. The Gaussian fitting of the bins represented by red line in the panel.
\label{Plx_Hists}}
\end{figure}

According to \cite{lindegren2021gaia}, the parallax measurements of the Gaia project are biased due to problems in the instrument and data processing. 
The authors found the biases in the parallax measurements of the observed sources depend on their $G$ magnitudes, $(G_{BP}-G_{RP})$ colours, ecliptic latitudes, and  the effective wavelength and the pseudo-colour of the observation. \cite{lindegren2021gaia} formulated two functions to compute the biases in the measured parallaxes of the sources included in the Gaia DR3 with either five-parameter solutions or six-parameter solutions ($Z_5$ \& $Z_6$).  The corrected parallaxes are computed by $\varpi_i^{corr}=\varpi_i - Z(x_i)$. 
They found that the parallax biases of the sources having five-parameter solutions vary between $-$94 and 36  $\rm \mu as$, and those of the sources having six-parameter solutions vary between $-$151 and 130  $\rm \mu as$. 
Python implementations of the two functions (\textit{gaiadr3-zeropoint} python package) available in the Gaia web pages \footnote{\url{https://www.cosmos.esa.int/web/gaia/edr3-code}} \citep{lindegren2021gaia} which were used in this study to calculate the parallax biases of the selected probable members. 
We found that the mean parallax zero-point of the probable members is equal to $-$26.2  $\rm \mu as$. The corrected  parallax measurements is equal to 0.3122 $\pm$ 0.0351 mas, as shown in Figure \ref{Plx_Hists}, its corresponding distance (distance modulus) value is 3.2031 $\pm$ 0.3601 kpc (12.20 $\pm$ 0.26 mag).

Our results are in a quite fair agreement with \cite{monteiro2020fundamental} ( which was based on  Gaia DR2 data). However, due to the use of  the more precise Gaia DR3 data, we obtained a more precise estimates of the cluster's proper motion and parallax.  
%

%
%------------------------------------------------------------------------------------------------------------------------------------------------
%------------------------------------------------------------------------------------------------------------------------------------------------
\section{Colour-magnitude diagram, Age and  Distance}\label{CMD-Age-Dist}

\begin{figure}%[htp]
\captionsetup[subfigure]{labelformat=empty} % to remove subfigure labels
\centering
\subfloat[]{\scalebox{.5}{\includegraphics{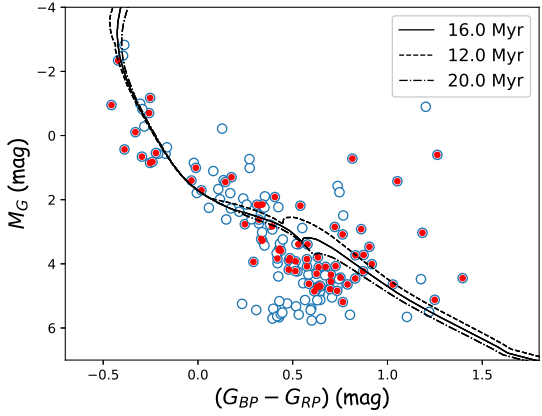}}}

\caption{The $G$ versus $G_{BP} \,- \,G_{RP}$ CMD, where open circles mark all stars that lay in the cluster zone while red-filed circles represent members having $P_\mu \geq 80\,\%$. 
\label{Gaia-cmds}}
\end{figure}

The most fundamental parameters of a cluster such as age, distance, reddening and metallicity can be simultaneously determined by fitting  theoretical isochrones to the produced Colour-magnitude diagram (CMD). In the current work, we used \cite{marigo2017new} theoretical isochrones with metallicity, Z=0.0152.
Using the member stars , we plotted the {\it G } versus $(G_{BP} - G_{RP})$ CMD as shown in Figure \ref{Gaia-cmds}. 
After comparing the observed data with different isochrones of different ages and by visual inspection, the best fitting isochrone is 16.00 Myr and the estimated value of the observed distance modulus of the cluster is of 15.00$\pm$0.30 mag.
The $ E(G_{BP}-G_{RP})$ colour excess ranges between 1.15 and 1.35 mag, so we adopted the average value as 1.25 mag, its $E(B-V)$ colour excess can be computed by $E(B-V)$=0.775$ \times E(G_{BP}-G_{RP})$ \citep{cardelli1989relationship} which is equal to 0.97 mag. The line-of-sight extinction coefficient is computed using $ A_G=2.74 \times E(B-V)$ \citep{casagrande2018use,zhong2019substructure}.Thus, the visual extinction coefficient is found to be  2.66 mag. 
The distance of the cluster can be obtained using its measured distance modulus by the following relation
\begin{equation}
d=10^{((m-M)_{obs}-A_{G}+5)/5}\,.
\end{equation}
The distance of Stock 3 after correcting the ISM extinction is 2945.3 $\pm$ 699.8 pc, which agrees with the measured parallax-based distance within uncertainties. Moreover, Our estimated age of Stock 3 agrees well with that of \cite{monteiro2020fundamental} (Age= 16.83 $\pm$ 3.40 Myr).

\begin{figure}[htp] 
\captionsetup[subfigure]{labelformat=empty} % to remove subfigure labels
\centering
\subfloat[]{\scalebox{.5}{\includegraphics{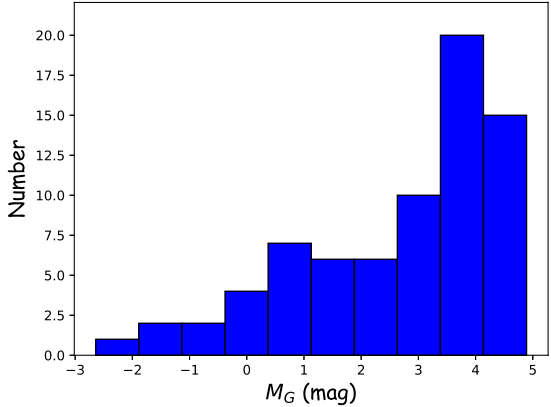}}}

\caption{The luminosity function with interval bins of 1.00 mag. (of stars with probability $\geq$ 80\%)
\label{LFs}}
\end{figure}

Using the measured parallax-based distance, the distance of the cluster from the Galactic plane, $Z_\odot$, and its projected distances from the Sun, $X_\odot$ and $Y_\odot$, and its distances from the Galactic centre $R_{GC}$ were calculated using the following equations :
\begin {align}
X_{\odot} = d \cos b \cos l\,, Y_{\odot} = d \cos b \sin l \,,  \nonumber\\
&\hspace{-5.0cm}Z_{\odot} = d \sin b \, \&
\end{align}
\begin{equation}
R_{GC} = \sqrt {R_{\odot}^{2}+(d \cos b)^{2}-2R_{\odot}^{2}d \cos b \cos l}\,, 
\end{equation}
where $d$ is the distance to the Sun in parsec, $R_\odot $ is the distance between the sun and the Galaxy centre. According to \cite{bland2019galah}, $R_\odot $=  8.20 $\pm$ 0.10 kpc.

 The values of $X_\odot$, $Y_\odot$ and $Z_\odot$ are equal to $-$1.85 kpc, 2.61 kpc, and $-$0.03 kpc ($\pm$ 0.20 kpc), respectively. Hence, its distances to the Galactic centre $(R_{GC})$ , is 10.39 $\pm$ 0.10 kpc.
\section{Luminosity and Mass Functions}\label{LM-MF-sect}  
After determining the probable members, it became easy to get the luminosity function (LF) and mass function (MF) of the cluster. The estimated distance modulus value can be used to compute the absolute magnitudes of the cluster members, $ M_{G}$, using their observed \textit{G} magnitudes. 
Figure \ref{LFs} shows the LF which reflects the incompleteness in the photometry in the ranges of $ M_{G}$ from 1 to 3 mag and greater than $\approx$ 4 mag.

Afterwards, by using the data of the selected isochrones from \cite{marigo2017new}, we  constructed the mass luminosity relations (MLR) by fitting the isochronous mass and luminosity data with a fourth-degree polynomial over an absolute magnitude interval between $-4.00$ and 5.00 mag. This interval  includes the estimated absolute magnitudes of the member stars. The masses of the probable members range between 1.00 and 9.50 $M_\odot$.
Consequently, that polynomial equation was used to get the mass estimates of the probable members. The MLR calculated as follows: 
\begin {align}
M_{Stock 3} = 3.930-1.587\; M_{G}+ 0.262\; M_{G}^{2} -  \nonumber\\
&\hspace{-5.0cm}0.013\; M_{G}^{3} - 0.005\;M_{G}^{4}
\end{align}

Table \ref{Stock3_members_list} provides a list of the estimated masses and the probabilities for a sample of the probable members. The complete list of the cluster is available online. Our mass estimates of the member stars show that Stock 3 includes 9 B-type stars and 64 later type stars.

\begin{figure} %[htp]
\captionsetup[subfigure]{labelformat=empty} % to remove subfigure labels
\centering
\subfloat[]{\scalebox{.45}{\includegraphics{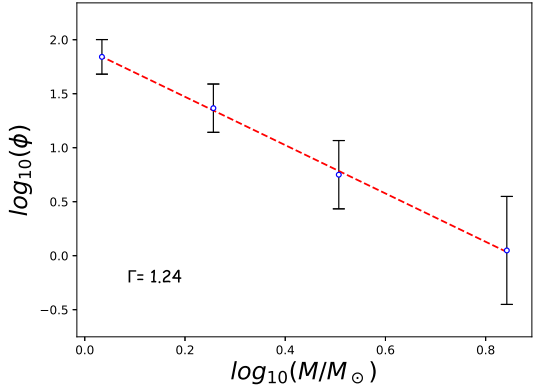}}}

\caption{The MF of the cluster. In the figure, $\phi$ represents $\frac{dN}{d M}$, the error bars represent the statistical errors computed by $1/\sqrt{N}$, and the dashed, red line illustrates a least square fit to the entire mass range. 
\label{IMFs}}
\end{figure}

\cite{salpeter1955luminosity} found that the number of stars decreases with the increase of stellar mass in a power-law relation of a power law index ($\alpha$) equal to 2.35. The slope of the MF of  a specific cluster can be derived using the following equation of \cite{bisht2020investigation},
%After dividing the obtained masses into bins and counting the number of stars in each bin of mass, we applied linear fitting to determine the slope of mass distribution as given by \citep{Bis20}, which represent the IMF slope that is obtained from equation (\ref{IMF_eq}): 
%
\begin{equation}
%\frac{dN}{dM}\propto M^{-\alpha} \,,
\log \frac{dN}{d M}=-(1+\Gamma) \times \log(M)+ constant.
\label{MF_eq}
\end{equation}
In equation \ref{MF_eq}, Salpeter's power law index $\alpha$ is equal to $1+\Gamma$. The total mass of the cluster ($M_{cl}$) in units of solar mass was estimated by summing up the masses of all cluster members which found to be equal to 135.54 $\pm$ 15.13 $M_\odot$, where errors are attributed to the uncertainties of distance modulus and age.
The slope of the MF ($\Gamma$), is equal to 1.24, as shown in Figure \ref{IMFs}. This value is close to the Salpeter's value ($\Gamma$+1= 2.35) for the mass range 0.40 $\leq M⁄ M_\odot< $  10.00 \citep{salpeter1955luminosity}. 

%\begin{landscape}
\begin{table*}%[htbp]
  \centering
\caption{The estimated probabilities and masses of a sample of the selected members  in addition to their astrometric and photometric data obtained from Gaia DR3. The complete table is available in the electronic form.
\label{Stock3_members_list}}
\scalebox{0.8}{
% \begin{tabular}  {| l | l |  l |  l |  l | r | r |  r | r | r | r | | }
\begin{tabular}{lcccccccccc}
 \hline  \hline
%&&&&&&&&&&\\  
 ID &  $\alpha$ (2000)  &   $\delta$ (2000)   &   $\varpi_{cor}$           & $\mu_\alpha \cos \delta$ &   $\mu_\delta$   &   $G$   &  $G_{RB}-G_{PB}$      &   Prob     & Mass  \\
& (deg) & (deg) & (mas)  & ($\rm mas\;yr ^{-1 } $) & ($\rm mas\;yr ^{-1 } $)  & (mag) & (mag) & ($\%$) & $\rm M_\odot$ \\ 
\hline\\    
M  1 & 18.033360 & 62.188053 & 0.3429 $\pm$  0.0279 & $-$2.114 $\pm$   0.022 & $-$0.534 $\pm$   0.029 &15.433 &0.863  &  99 &  3.29 $\pm$  0.42 \\
M  2 & 18.001565 & 62.292879 & 0.3692 $\pm$  0.0212 & $-$2.139 $\pm$   0.018 & $-$0.482 $\pm$   0.023 &14.895 &0.920  &  99 &   4.10 $\pm$  0.52 \\
M  3 & 18.190841 & 62.220717 & 0.3940 $\pm$  0.0341 & $-$2.117 $\pm$   0.028 & $-$0.468 $\pm$   0.036 &16.007 &1.239  &  99 &   2.61 $\pm$  0.32 \\
M  4 & 17.999354 & 62.296598 & 0.4147 $\pm$  0.0317 & $-$2.148 $\pm$   0.027 & $-$0.524 $\pm$   0.035 &15.819 &1.010  &  99 &   2.81 $\pm$  0.35 \\
M  5 & 18.012249 & 62.262776 & 0.3350 $\pm$  0.0151 & $-$2.150 $\pm$   0.012 & $-$0.572 $\pm$   0.016 &13.825 &0.998  &  99 &   6.12 $\pm$  0.78 \\
&&\multicolumn{7}{c}{...........................................................................................................................}&&\\
&&&&&&&&&&\\\hline  
 \end{tabular}}
\end{table*}

\begin{figure}  %[htp]
\captionsetup[subfigure]{labelformat=empty} % to remove subfigure labels
\centering
\subfloat[]{\scalebox{.45}{\includegraphics{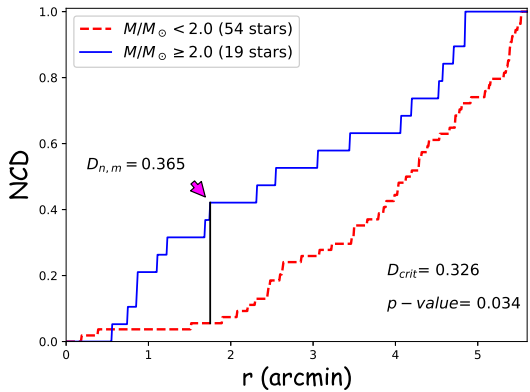}}}
\caption{The cumulative radial distribution of stars in two different mass ranges, M$<$2.00 $M_\odot$ and M$\geq$2.00 $M_\odot$ of Stock 3.
\label{CRDs}}
\end{figure}

%------------------------------------------------------------------------------------------------------------------------------------------------
%------------------------------------------------------------------------------------------------------------------------------------------------
\section{ Dynamical State of The Cluster}\label{dyan_state_sect}              

Many observed star clusters show mass segregation  throughout the distribution of the stars with distance from the centre, where the massive stars lie closer to the centre, while the low mass ones lie closer to the outer edge of the cluster. It is still a matter of debate whether this observed mass segregation results from the dynamical interactions of the stars within the clusters or the cluster's stars were formed that way. 
The search of the present mass segregation can be done by constructing the cumulative radial distribution of member stars within different mass ranges and by looking for differences between these distributions. Next, we computed the relaxation time scale of the cluster. If the cluster’s age is less than the relaxation time and it shows a clear mass segregation in its structure, this indicates that the cluster was born this way.  On the other hand, if the cluster’s age is less than the relaxation time and there is no sign of mass segregation in its structure, this concludes that mass segregation results from dynamical interactions.  If the cluster's age is larger than the  relaxation time, we can not resolve the debate.
For this purpose, we used to the Kolmogorov–Smirnov test to look for mass segregation. As it searches for the maximum difference between the cumulative radial distributions of two samples with sizes n and m, $D_{n,m}$, which is computed using the following relation:
\begin{equation}
D_{n,m}=sup_x |F(x)-G(x)|
\end{equation}
The null hypothesis which is the two distributions are similar is rejected, if $D_{n,m}$ is greater than the critical value  $D_{crit}$ computed for the same sizes as those of the two samples and for a specific level of significance $\alpha$.   
Figure \ref{CRDs} presents the cumulative radial distribution of stars in two different mass ranges for the cluster, $M\, < \,2.00 \,M_\odot$ and $ M\geq 2.00 \,M_\odot$. It shows that the two radial distribution of stars of Stock 3 in the two mass ranges are statistically different at 10\% significance level. The probability of resemblance, p-value, is shown in the figure which is equal to 3.4 \% for Stock 3.
In order to know whether the cluster is dynamically relaxed or not, we estimated the relaxation time ($T_R$)  which can defined as the time in which the stellar velocity distribution becomes Maxwellian and calculated using the following equation:

\begin{equation}  T_R= \frac {8.9 \times 10^{5}\sqrt {N} r_{h}^{1.5}}{\sqrt {<m>} \log (0.4 N)}\,,  \end{equation}

where N is the number of the probable member stars, $r_h$ is the half mass-radius in parsecs computed using \cite{Lar06} equation
\begin{equation} r_h=0.547 \times r_c \times \Big( \frac{r_t}{r_c}\Big)^{0.486}\,, \end{equation}
where $r_t$ is the tidal radius, would be defined in the next section, and $<m>$ is the average mass of the cluster’s members, in unit of solar mass, which can be calculated by the relation, $ <m> =M_{cl}⁄N$ , where $ M_{cl} $ is the cluster's mass, which gives $ <m>$ equal to 1.86 $M_\odot$ ⁄star.
The cluster is considered as a dynamically relaxed cluster if $\tau \,=\,Age/T_R \gg 1$, where $\tau$ is known as the dynamical evolution parameter. 
The estimated relaxation time for Stock 3 is 5.28 $\pm$ 0.94  Myr, which is close to its age. Therefore, the cluster is not dynamically relaxed yet. 
The cluster shows mass segregation in its internal structure, however its age is larger than its relaxation time and as consequence, such result can not resolve the debate of the mass segregation phenomenon. 

%$T_R$ (Myr)  & 12.55 $\pm$ 1.48  & 3.24 $\pm$ 0.36  \\ 
%------------------------------------------------------------------------------------------------------------------------------------------------
%------------------------------------------------------------------------------------------------------------------------------------------------
\section{Velocity Ellipsoid Parameters (VEPs) and the Coherent Point}\label{VEP_sec} 

\subsection{Ellipsoidal motion and the kinematical structure}

In the following, we estimated the evaporation time $\tau_{ev}$  ($\approx10^2 \,  T_R$), which is defined as the time needed to eject all member stars by internal stellar encounters \citep{adams2001modes}, low-mass stars continue to escape from the cluster, mainly at low speeds through Lagrange points \citep{kupper2008structure}. For a cluster to remain bound in face of the Galactic disruptive effects, the escaping velocity ($V_{esc}$) of rapid gas removal from the cluster must equal to ($V_{esc}=R_{gc} \times \sqrt{\frac{2GM_{cl}}{3 r_t^3} }$) which is as a function of the cluster's total mass and tidal radius \citep{fich1991mass,fukushige2000time}. The tidal radius is defined as the distance at which two opposite forces are neutralized, one is due to cluster's own gravity and the other is devoted to the gravitational attraction toward the Galactic centre. Hence, the tidal radius can be computed  by applying the equation of \cite{jeffries2001photometry} which can be expressed as:
\begin{equation}
r_t=1.46 M_{cl}^{1/3}
\end{equation}
The tidal radius+ is found to be 7.50 $\pm$ 0.28 pc. 
To highlight the gravitationally bound system of the stellar groups in a limited volume of space within the Galactic system characterized by the parallelism and equality of their motions, we studied the velocity ellipsoid parameters VEPs and using a computational algorithm developed by \citet{elsanhoury2018pleiades, bisht2020investigation, elsanhoury2021photometric, bisht2021detailed, bisht2022comprehensive}. 
Member stars have common line-of-sight velocities, $V_r$. The line-of-sight velocity of Stock 3 is equal to $-6.00$ (km $\rm s^{-1}$) \citep{kharchenko2013global}.
In this manner, the velocity components ($V_X$, $V_Y$, $V_Z$; $\rm km\ s^{-1}$) along x, y and z-axes in the coordinate system with respect to the Sun were computed using Eq.  (9) of \cite{elsanhoury2022comprehensive}. The calculations in this section was done using only members with positive parallaxes.

\begin{figure} %[htp]
\centering
\subfloat[]{\scalebox{.5}{\includegraphics{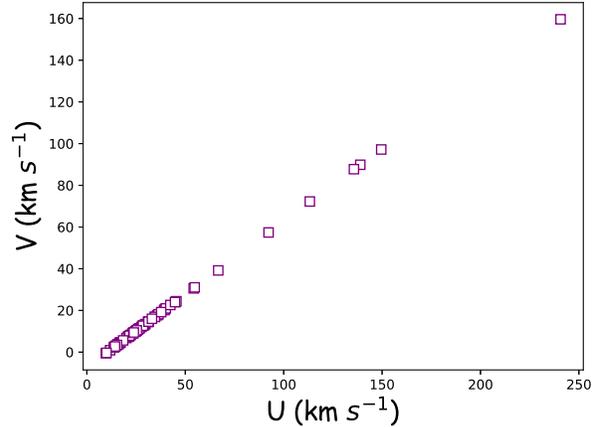}}}\\
\subfloat[]{\scalebox{.5}{\includegraphics{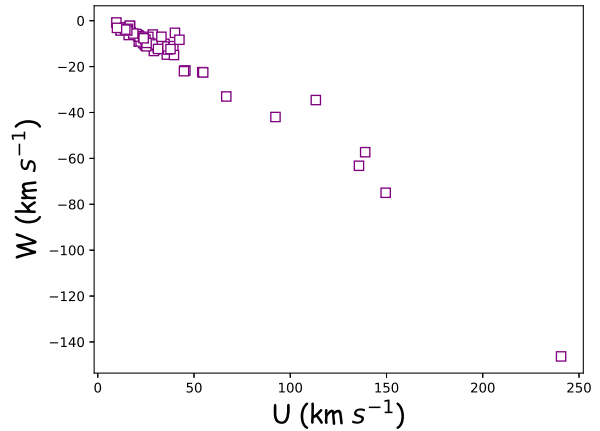}}}
\caption{The ($UVW$; km $\rm s^{-1}$) plots of spatial space velocity components along the Galactic coordinates of the extended Stock 3 member stars. $U$ is directed to the anti-centre of the Galaxy, $V$ is in the direction of rotation of Galaxy, and $W$ towards the north pole of the Galaxy. 
\label{UVW_Fig}}
\end{figure}

\begin{figure} %[htp]
\captionsetup[subfigure]{labelformat=empty} % to remove subfigure labels
\centering
\subfloat[]{\scalebox{.5}{\includegraphics{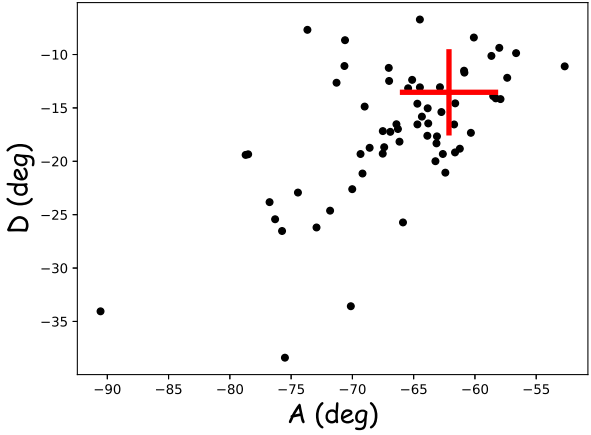}}}

\caption{The $AD$ – diagram. The red plus cross-matched symbol indicates to the apex position ($A$, $D$). 
\label{AD-diagram}}
\end{figure}

The spatial space velocity components ($U$, $V$ \& $W$; $\rm km\;s^{-1}$) of member stars on the celestial sphere along with Galactic coordinates were calculated using Eq. (10) of \cite{elsanhoury2022comprehensive}. The distributions of these space velocities are shown in Figure \ref{UVW_Fig}. It can be inferred from the figure that the open cluster Stock 3 has an extended velocity in ($UVW$). We can calculate the apex position; into which member stars will coherently be directed towards a point ($A$, $D$) defined as convergent point (or apex) by intersection of vectors of spatial velocities of stars with the celestial sphere. We adopted the $AD$ – diagrams method given by \cite{chupina2001geometry, chupina2006kinematic}, where they have used the notion of an individual stellar apexes in equatorial coordinates for the members through space velocity vectors ($V_x$, $V_y$, $V_z$ ) as follows:
\begin{equation}
A=\tan^{-1} \left(\frac{\overline{V}_y}{\overline{V}_x}\right)\, \& \, D=\tan^{-1} \left( \frac{\overline{V}_z}{\sqrt{\overline{V}_x^2+\overline{V}_y^2}}\right)
\end{equation}
Figure  \ref{AD-diagram} gives the apex equatorial coordinates ($A$, $D$) for Stock 3 open cluster, our numerical results along with VEP are given in Table \ref{dyn_evol_paras}.

\subsection{Other kinematic structure parameters}

\subsubsection{The centre of the cluster ($x_c$, $y_c$, $z_c$)}

For $N_i$ member stars located at distance $d_i$ (pc) coordinated with ($\alpha_i$, $\delta_i$), the cluster center ($x_c$, $y_c$, $z_c$) in units of pc were estimated using Eq.  (12) of \cite{elsanhoury2022comprehensive}.

\subsubsection{The Solar elements} 

The components of the Sun's velocities ($U_\odot$, $V_\odot$, and $W_\odot$) are given like, ($U_\odot$= $-\overline{U}$), ($V_\odot$= $-\overline{V}$), and ($W_\odot$= $-\overline{W}$ ).
\begin{equation}
S_\odot=\sqrt{\overline{U}^2 +\overline{V}^2+\overline{W}^2}\,,
\end{equation}
and the position of the Solar apex in Galactic coordinates are ($l_A$,$b_A$) as function of spatial space velocity.
\begin{equation}
l_A=\tan^{-1}\left(\frac{-\overline{V}}{\overline{U}} \right)\, \& \, b_A=\sin^{-1}\left(\frac{-\overline{W}}{S_\odot} \right).
\end{equation}
Now consider the position along x, y, and z-axes in the coordinate system whose centred at the Sun, then the Sun's velocities ($X_\odot^\bullet$, $Y_\odot^\bullet$, $Z_\odot^\bullet$ ) are given with ($X_\odot^\bullet$= $-\overline{V}_x$ ),($Y_\odot^\bullet$= $-\overline{V}_y$ ), and ($Z_\odot^\bullet$= $-\overline{V}_z$). Similarly, the Solar elements with radial velocities.

\begin{equation}
S_\odot=\sqrt{(X_\odot^\bullet)^2+(Y_\odot^\bullet)^2+(Z_\odot^\bullet)^2}\,,
\end{equation}
\begin{align}
\alpha_A=\tan^{-1} \left( \frac{Y_\odot^\bullet}{X_\odot^\bullet} \right) \, \& \nonumber \\ 
\delta_A=\tan^{-1} \left(\frac{Z_\odot^\bullet}{\sqrt{(X_\odot^\bullet)^2+(Y_\odot^\bullet)^2}}\right) 
\end{align}
where ($\alpha_A$, $\delta_A$) are the Galactic right ascension and declination of the Solar apex and $S_\odot$ is considered as the absolute value of the Solar velocity relative to the group under study.

\subsubsection{Stock 3 Morphology with 3D} 

\begin{figure} %[htp]
\captionsetup[subfigure]{labelformat=empty} % to remove subfigure labels
\centering
\subfloat[]{\scalebox{.5}{\includegraphics{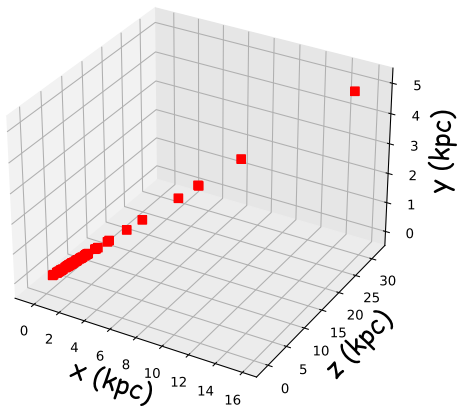}}}	
\caption{The 3D spatial morphology plots in heliocentric Cartesian coordinates (X, Y, Z; kpc) of Stock 3. 
\label{3D-plots}}
\end{figure}

We analyse the 3D spatial position of member stars of the cluster in heliocentric Cartesian coordinates ($X$, $Y$, $Z$; pc).
\begin{equation}
X=d \cos \delta \cos \alpha ,\, Y=d \cos \delta \sin \alpha                                   \, \&\, Z=d \sin \delta
\end{equation}
The 3D morphology for this cluster was plotted as shown in Figure \ref{3D-plots}, and it is noticeable that the stars of the cluster expand through elongated regions in space; especially in the $XZ$-plane. This expansion may be considered as fast gas expulsion and virilization \citep{pang2021disruption}.

\begin{table}%[htbp]
  \centering
\caption{Our obtained dynamical evolution parameters with different times and the VEPs with their kinematics.
\label{dyn_evol_paras}}
\scalebox{1.0}{
% \begin{tabular}  {| l | l |  l |  l |  l | r | r |  r | r | r | r | | }
\begin{tabular} {lc}
 \hline  \hline
%&&&&&&&&&&\\  
  Parameters  &   Stock 3   \\
\hline
$T_{R}$ (Myr)	& 	5.28 $\pm$ 0.94 \\
$\tau_{ev}$ (Myr) &	528 $\pm$ 94     \\
$\tau$ &				3.03	 \\
$V_{esc}$ (km $\rm s^{-1}$) & 	315.29 $\pm$ 17.76		\\
$\overline{V}_x$  (km $\rm s^{-1}$) & 21.04 $\pm$ 4.59   \\
$\overline{V}_y$  (km $\rm s^{-1}$) & $-$39.76 $\pm$ 6.31   \\
$\overline{V}_z$  (km $\rm s^{-1}$) & $-$10.83 $\pm$ 3.29   \\
$A\,(^\circ)$ & $-$62.12 $\pm$ 0.13   \\
$D\,(^\circ)$ & $-$13.54 $\pm$ 0.28   \\
$\overline{U}$  (km $\rm s^{-1}$) & 38.85 $\pm$ 6.23   \\
$\overline{V}$  (km $\rm s^{-1}$) & 19.86 $\pm$ 4.46   \\
$\overline{W}$  (km $\rm s^{-1}$) & $-$15.37 $\pm$ 3.92   \\
$x_c$ (pc) & 2056.30 $\pm$ 45.35  \\
$y_c$ (pc) & 669.13 $\pm$ 25.87   \\
$z_c$ (pc) & 4111.43 $\pm$ 64.12   \\
$S_\odot$  (km $\rm s^{-1}$) & 46.26 $\pm$ 6.8  \\
($l_A$, $b_A$)$^\circ$ & $-$27.08 , 19.41   \\
($\alpha_A$, $\delta_A$)$^\circ$ & $-$62.12 , 13.54  \\
\hline  
\end{tabular}}
\end{table}

\section{CONCLUSION}\label{conc-sect}

  In present work, we derived photometric and astrometric parameters of the open star cluster Stock 3 using
 Gaia DR3. We determined  73 most probable member stars with membership probabilities $P_\mu$ $\geq$ 80 $\%$. We utilized these Gaia-based probable members to derive all the parameters.  Our results are quite compatible with the estimated parameters obtained in some of the most recent previous studies as shown in Table \ref{Stock3_results}. The major conclusions of the present investigations can be summed up as follows:

    The coordinates of the cluster's centre are:
\begin{itemize}
\item  $\alpha =01^{h} 12^{m} 9^{s}.40$ \&  $\delta= +62^{\circ} 15' 23.08''$ 

\end{itemize}

The radius of the cluster was evaluated as 5.60 $\pm$ 0.43 arcmin. The radius represents the distance from the cluster's centre where the cluster's density melted with the background density.

Using the most probable members of the cluster identified using the Gaia DR3 proper motion data,  we constructed the proper motion and parallax histograms of these Gaia-based probable members. Here are what we calculated:
\begin{itemize}
	\item $\mu _{ \alpha }\cos \delta$ = $-$2.0673 $\pm$ 0.0173 
	\item $\mu_\delta$ = 0.4657 $\pm$ 0.0192 $\rm( mas\;yr^{-1})$
	\item $\varpi_{corr}$ = 0.3122 $\pm$ 0.0351 (mas) (corresponding distance= 3.2031 $\pm$ 0.3601 kpc) 
\end{itemize}

   Using Gaia DR3, We constructed the CMD and fitted it with the theoretical isochrones of \cite{marigo2017new}, determining its age as 16.00 $\pm$ 4.00  Myr. Its isochrone-based distance of Gaia photometry is 2945.3 $\pm$ 699.8 (pc) which is close to the parallax-based distance. 

Next, we derived the distance of the cluster from the Galactic plane, $Z_{\odot}$, its projected distances from the Sun, $ X_{\odot}$, $Y_{\odot}$, and the distance from the Galactic centre $R_{ GC }$, all are listed in Table \ref{all_results}.

The total mass was obtained and the slope of the IMF which considered to be in a fair agreement with \cite{salpeter1955luminosity} value. 
\begin{itemize}
	\item $M\textsubscript{Stock 3}$= 135.54 $\pm$ 15.13 $M_{\odot}$ \&   $\alpha$= 2.24 $\pm$ 0.69
\end{itemize}

     The relaxation time calculation of Stock 3 was computed which shows that it is not dynamically relaxed cluster and it shows a statistically, significant mass segregation. The observed mass segregation might be attributed to the dynamical interactions between the cluster members.

Finally, we provided estimates of the spatial space velocities ($UVW$) which are used to get estimates of the Solar elements like $S_\odot$ that is equal to 46.26 (km $\rm s^{-1})$ for Stock 3.

Moreover, the kinematical parameters can be deduced with the VEP, into which the vertex position was estimated using the AD-diagram method that exists at $-62^\circ .12$ \& $-13^\circ .54$ for Stock 3. On the other hand, the spatial velocities in space ($V_x$, $V_y$, $V_z$) presented the spatial morphology of the cluster.

\begin{center}
\begin{table}[htbp]
\centering
\caption{Summary of the results obtained for the Stock 3 in the present work.
\label{all_results}}
\scalebox{0.85}{
\begin{tabular}{|l |c |} 
 \hline
 Parameter & Stock 3  \\
\hline
 $\alpha $  (deg) & 18.039268     \\  
$\delta $ (deg)  &  62.25641   \\%  &&\\
  $\ell \;\&\;b$ (deg)  &125.335, $-$0.520  \\ % &&\\
r (arcmin)   & 5.60 $\pm$ 0.43  \\ % &&\\
$ (m-M)_{obs}$ (mag) & 15.0$\pm$0.3  \\%  &&\\
$ E( G_{BP}-G_{RP})$ (mag) & 1.25 $\pm$ 0.1\\%  &&\\
 $E(B-V)$ (mag) &0.97 $\pm$ 0.1  \\ %&&\\
$ A_{G}$ (mag) & 2.66 $\pm$ 0.24  \\%  &&\\
% d($m-M$) (kpc) &  2.8799 $\pm$ 0.5860 & 2.9878 $\pm$ 0.6090  \\%  &&\\
 d($m-M$) (kpc) &  2.945 $\pm$ 0.700  \\%  &&\\
Age (Myr) & 16.00 $\pm$ 4.00   \\ %  &&\\
$\varpi$ (mas) & 0.3122 $\pm$ 0.0351    \\ 
d($\varpi$)  (kpc) &  3.2031 $\pm$ 0.3601 \\%  &&\\ 
$\mu _{ \alpha }\cos \delta \; (mas\;yr^{-1})$  &$-$2.0673 $\pm$ 0.0173    \\  
$\mu_\delta \; (mas \;yr^{-1})$ & $-$0.4657 $\pm$ 0.0192   \\%  &&\\
$N_{stars}$  & 73  \\%  &&\\
$X_{\odot}$ (kpc) &  $-$1.85 $\pm$ 0.20  \\ %  &&\\
$Y_{\odot}$ (kpc) & 2.61 $\pm$ 0.20   \\%  &&\\
$Z_{\odot}$ (kpc) & $-$0.03 $\pm$ 0.20 \\%  &&\\
$R_{GC}$(kpc) & 10.39 $\pm$ 0.10   \\ %  &&\\
$M_{cluster}$ ($M_{\odot}$)  &  135.54 $\pm$ 15.13    \\%  &&\\
$\alpha$ & 2.24 $\pm$ 0.69 \\  %  &&\\
%  $T_{relax}$ (Myr) & 12.55 $\pm$ 1.48  & 3.24 $\pm$ 0.36 \\ 
 \hline    
\end{tabular}}
\end{table}
\end{center}

\section*{Acknowledgement}
%We thank the anonymous referee for his/her valuable comments and suggestions. 
%
This work has made use of  data from the European Space Agency (ESA) mission {\it Gaia} (\url{https://www.cosmos.esa.int/gaia}), and processed by the {\it Gaia} Data Processing and Analysis Consortium (DPAC, \url{https://www.cosmos.esa.int/web/gaia/dpac/consortium}). Funding for the DPAC has been provided by the national institutions, in particular the institutions participating in the  {\it Gaia}  Multilateral Agreement. The authors extend their appreciation to the Deanship of Scientific Research at Northern Border University, Arar, KSA for funding this research work through the project number ‘‘NBU-FFR-2025-237-01”.

%\section*{FUNDING}
%This work was supported by the\ldots

\section*{CONFLICT OF INTEREST}
The authors declare no conflicts of interest.

\end{document}

%% file: sao_cmd_author.tex
\def\squareforqed{\hbox{\rlap{$\sqcap$}$\sqcup$}}

\def\sq{\ifmmode\squareforqed\else{\unskip\nobreak\hfil
\penalty50\hskip1em\null\nobreak\hfil\squareforqed
\parfillskip=0pt\finalhyphendemerits=0\endgraf}\fi}

\def\utw{\smash{\rlap{\lower5pt\hbox{$\sim$}}}}

\def\udtw{\smash{\rlap{\lower6pt\hbox{$\approx$}}}}

\def\diameter{{\ifmmode\mathchoice
{\ooalign{\hfil\hbox{$\displaystyle/$}\hfil\crcr
{\hbox{$\displaystyle\mathchar"20D$}}}}
{\ooalign{\hfil\hbox{$\textstyle/$}\hfil\crcr
{\hbox{$\textstyle\mathchar"20D$}}}}
{\ooalign{\hfil\hbox{$\scriptstyle/$}\hfil\crcr
{\hbox{$\scriptstyle\mathchar"20D$}}}}
{\ooalign{\hfil\hbox{$\scriptscriptstyle/$}\hfil\crcr
{\hbox{$\scriptscriptstyle\mathchar"20D$}}}}
\else{\ooalign{\hfil/\hfil\crcr\mathhexbox20D}}%
\fi}}